\documentclass[prl, aps, reprint]{revtex4-1}
\usepackage[utf8]{inputenc}
\usepackage{amsmath,amssymb,amsfonts,ulem}
\usepackage{graphicx}
\usepackage{upgreek}
\usepackage[usenames, dvipsnames]{color}

\definecolor{JB}{rgb}{1, 0.6, 0.}
\definecolor{PG}{rgb}{1.0, 0.0, 1.0}

\begin{document}

\title{Loss of ultracold $^{87}$Rb$^{133}$Cs molecules via optical \\ excitation of long-lived two-body collision complexes}

\author{Philip D. Gregory}
\author{Jacob A. Blackmore}
\author{Sarah L. Bromley}
\author{Simon L. Cornish}
\affiliation{Department of Physics, Joint Quantum Centre (JQC) Durham-Newcastle, Durham University, South Road, DH1 3LE, United Kingdom}

\begin{abstract}
    We show that the lifetime of ultracold ground-state $^{87}$Rb$^{133}$Cs molecules in an optical trap is limited by fast optical excitation of long-lived two-body collision complexes. We partially suppress this loss mechanism by applying square-wave modulation to the trap intensity, such that the molecules spend 75\% of each modulation cycle in the dark. By varying the modulation frequency, we show that the lifetime of the collision complex is $0.53\pm0.06$\,ms in the dark. We find that the rate of optical excitation of the collision complex is \mbox{$3^{+4}_{-2}\times10^{3}$\,W$^{-1}$\,cm$^{2}$\,s$^{-1}$} for $\lambda=1550$\,nm, leading to a lifetime of $<100$~ns for typical trap intensities. These results explain the two-body loss observed in experiments on nonreactive bialkali molecules.
\end{abstract}

\maketitle
There is currently rapid experimental progress in the field of ultracold molecules~\cite{Danzl:2008, Ni:2008, Lang:2008, Takekoshi:2014, Molony:2014, Park:2015, Prehn:2016, Guo:2016, Rvachov:2017, Truppe:2017, Seesselberg:2018, Anderegg:2018, Collopy:2018}, spurred on by a range of exciting applications including the study of dipolar gases~\cite{Santos:2000, Ni:2009, Baranov:2012, DeMarco:2019}, quantum simulation~\cite{Micheli:2006, barnett:2006, Gorshkov:2011, Hazzard:2013, Sundar:2018, Blackmore:2018}, quantum computation~\cite{DeMille:2002, Ni:2018, Sawant:2020, Hughes:2019}, precision measurement~\cite{Flambaum:2007, Isaev:2010, Hudson:2011, Baron:2014, Andreev:2018}, and quantum-state controlled chemistry~\cite{Krems:2008, Bell:2009, Dulieu:2011, Balakrishnan:2016}. Densities in experiments are now sufficiently high that collisions between molecules are important and measurable. During collisions, pairs of molecules form 
transient collision complexes whose properties may affect the collision outcome. Such complexes are found throughout chemistry as intermediates in chemical reactions, but generally their ephemeral nature makes their detection challenging~\cite{Miller:1972, Bauer:1979}. However, at ultracold temperatures, collision complexes can be significantly longer lived, presenting a new opportunity to study their dynamics.

Collisions between ultracold heteronuclear molecules were first studied in fermionic $^{40}$K$^{87}$Rb~\cite{Ospelkaus:2010, Ni:2010}. Here, the lifetime is limited by reactive two-body collisions of the form $2\mathrm{KRb}\rightarrow\mathrm{K}_2 + \mathrm{Rb}_2$. The reactive nature of KRb collisions has been recently confirmed through direct detection of the intermediate complexes and reaction products~\cite{Hu:2020}. Thankfully not all bialkali molecules have energetically allowed two-body reactive collisions~\cite{Zuchowski:2013}, offering hope that stable molecular gases may be produced. However, experiments with nonreactive molecules such as bosonic $^{87}$Rb$^{133}$Cs~\cite{Takekoshi:2014, Gregory:2019} and $^{23}$Na$^{87}$Rb~\cite{Ye:2018, Guo:2018}, and fermionic $^{23}$Na$^{40}$K~\cite{Park:2015, Yang:2019}, have all observed fast losses from optical traps, characterised by two-body loss rates comparable to those found in the reactive case. Understanding the mechanism for this loss is of paramount importance to the development of the field.

Mayle~$et~al.$ proposed a possible mechanism for the loss of nonreactive molecules~\cite{Mayle:2012, Mayle:2013}. They argue that the large number of rovibrational states accessible in a collision will lead to a dense manifold of Feshbach resonances. Scattering in this highly resonant regime leads to the formation of long-lived collision complexes. Ordinarily, the complexes would simply break apart back into free molecules. However, in their proposal, the complex lifetimes are predicted to be sufficiently long that a further collision with a third molecule is possible, leading to loss of all three molecules from the trap~\cite{Mayle:2013}. Crucially, if the formation of collision complexes is the rate-limiting step, then the loss would appear to be two-body in nature, consistent with experimental observations~\cite{Ye:2018, Guo:2018, Gregory:2019, Yang:2019}.

The model of Mayle~$et~al.$~\cite{Mayle:2012, Mayle:2013} assumes that the lifetime of the complex $\tau_\mathrm{c}$ is related to the density of states $\rho$ by $\tau_\mathrm{c} = 2\pi\hbar\rho$. For collisions of RbCs in the rovibrational ground state, they predict a density of states of \mbox{$\rho/k_{\rm{B}} = 942\,\upmu$K$^{-1}$} and a lifetime of 45\,ms for the RbCs+RbCs complex (hereafter, (RbCs)$_2$)~\cite{Mayle:2013}. However, recent work indicates that the density of states was over-estimated by Mayle~$et~al.$, leading to complex lifetimes that are 2 to 3 orders of magnitude too large~\cite{Croft:2017, Christianen:2019_DOS}. Specifically, Christianen~$et$~$al.$ have estimated the rate of (NaK)$_2$+NaK complex-molecule collisions~\cite{Christianen:2019}, and find that for typical experimental densities, the lifetime associated with this rate is significantly longer than $\tau_\mathrm{c}$. They conclude that complex-molecule collisions are unlikely to be the cause of loss.

\begin{figure}[t]
    \centering
    \vspace{-0.7cm}
			\includegraphics[width=0.48\textwidth]{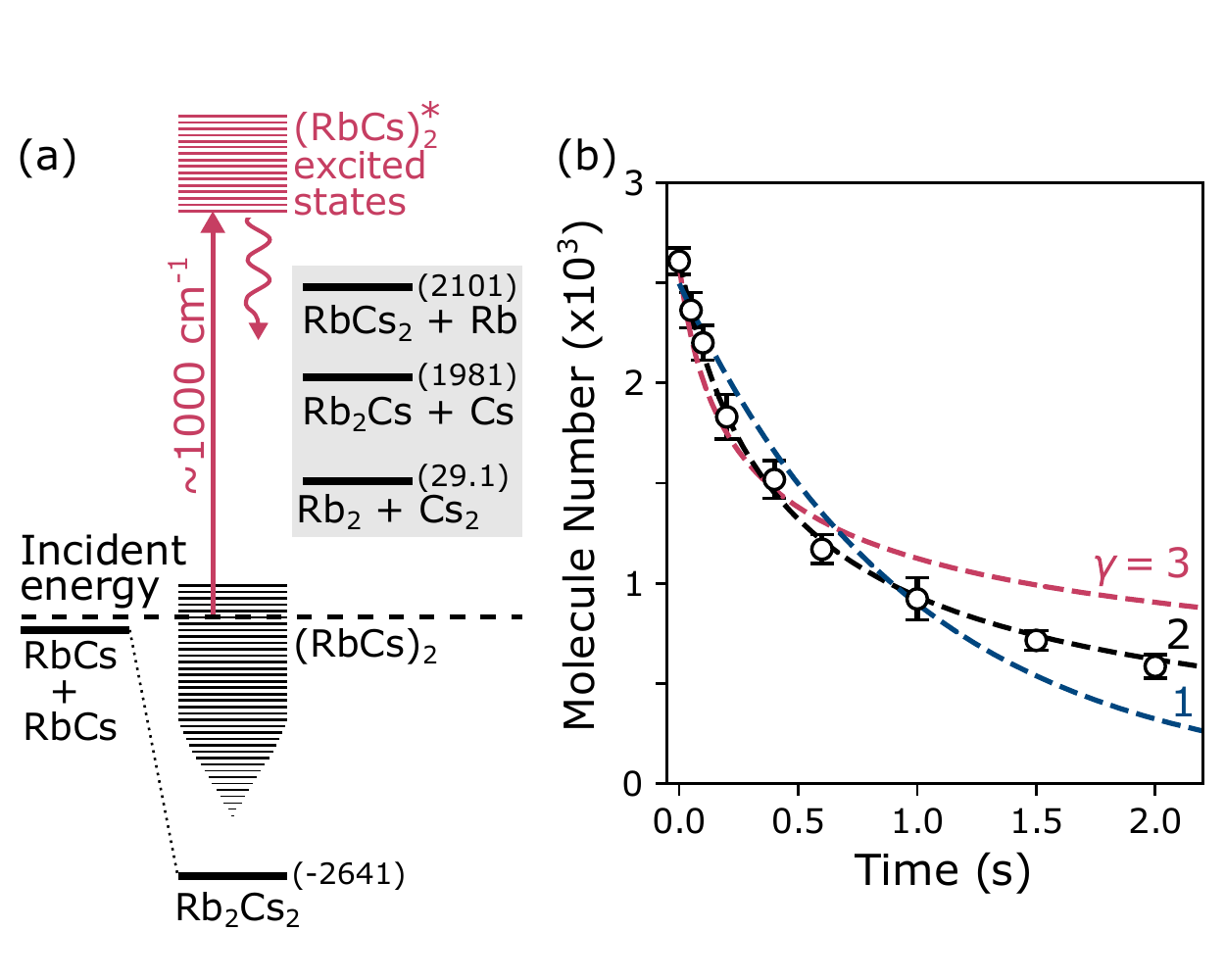}
	\vspace{-0.7cm}
    \caption{\label{fig:CWTrap} Two-body photoinduced loss of molecules. (a)~Energetics for loss of RbCs. Energy levels are labelled by their ground state energy (cm$^{-1}$), taken from~\cite{Zuchowski:2013, Byrd:2012}, with respect to the energy of two free RbCs molecules. All available atom-transfer reactions are energetically forbidden. However, the transient complex (RbCs)$_2$ can form due to a high density of states near the incident energy. These complexes may then absorb photons from the trap laser leading to the observation of two-body loss of RbCs. (b)~Collisional loss of ground-state RbCs in a CW 1064~nm trap. Dashed lines indicate models with first, second, and third order kinetics respectively.}
\end{figure}

Christianen~$et$~$al.$ instead propose that complexes may be removed via electronic excitation by the trapping light~\cite{Christianen:2019}, as depicted schematically in Fig.~\ref{fig:CWTrap}(a). They have performed calculations for the (NaK)$_2$ complex, and find laser excitation rates of $\sim1\,\upmu$s$^{-1}$. This is fast enough that essentially all complexes undergo laser excitation before they break apart. Again this loss mechanism manifests as a two-body process. Other bialkali complexes, including (RbCs)$_2$, are expected to have comparable electronic structure and therefore should exhibit similar excitation rates. In this case, optical excitation of complexes is expected to be the major cause of loss in ultracold gases of nonreactive molecules. 

In this letter, we show that fast optical excitation of collision complexes in an ultracold gas of $^{87}$Rb$^{133}$Cs is responsible for the loss of molecules. We partially suppress this loss mechanism by applying square-wave modulation to the optical trap intensity, such that the molecules spend 75\% of each modulation cycle in the dark. When the trap light is off, complexes can form and break apart without the risk of destructive optical excitation. Accordingly, a reduction in the loss rate is observed in the time-averaged trap, with the maximum fractional reduction simply equal to the duty cycle of the modulation. By studying the reduction in loss as a function of modulation frequency, we determine the  
lifetime of the (RbCs)$_2$ complex. By applying continuous wave (CW) light, in addition to the modulated trap light, we can probe the molecules in a low intensity environment during the dark periods of the trap, and hence measure the laser scattering rate for the complex. 

Our experiments use samples of approximately $3000$ molecules in their rovibrational and hyperfine ground state, at a temperature of 2.2\,$\upmu$K, and initial peak density of ${2\times10^{11}}$\,cm$^{-3}$. The methods used to produce the molecules are reported elsewhere~\cite{McCarron:2011, Koppinger:2014, Molony:2014, Gregory:2015, Molony:2016}. In this work, we confine the molecules in a 1064\,nm crossed optical dipole trap formed from a single laser beam aligned in a bow-tie configuration. The trap waists are 107(1)\,$\upmu$m and 74(1)\,$\upmu$m for the first and second passes, respectively. An acousto-optic modulator, placed between the two trap foci, shifts the laser frequency for the second pass by 80\,MHz to avoid interference effects. The intensity of the trap can be modulated using an optical chopper wheel, such that the molecules spend 75\% of each modulation cycle in the dark. When used, both beams are modulated with common phase at a frequency $f_{\mathrm{mod}}$ ranging from 400\,Hz to 5\,kHz. The trap is loaded by first preparing Feshbach molecules in a levitated 1550\,nm optical trap~\cite{Jenkin:2011,Koppinger:2014}. The molecules are subsequently transferred into the 1064\,nm trap via a 50\,ms linear intensity ramp, following which the 1550\,nm trap and magnetic levitation gradient are ramped off over 10 ms. Stimulated Raman Adiabatic Passage (STIRAP) is then used to transfer the molecules to the rovibrational and hyperfine ground state~\cite{ Molony:2014,Molony:2016}. Mechanical shutters block the $1550$\,nm light when it is not required. After a hold time in the 1064\,nm trap, we measure the number of molecules remaining by reversing the association process and detecting the constituent atoms using absorption imaging. We therefore only detect molecules remaining in the specific hyperfine state addressed in the STIRAP transfer.

We first report measurements of loss in a CW trap. Fig.~\ref{fig:CWTrap}(b) shows the number of molecules remaining in the 1064\,nm trap as a function of time, without intensity modulation. Dashed lines show fits to the results where we model the rate of change of density $n$ as $\dot{n}=-k_\gamma n^{\gamma}$, as described in~\cite{Gregory:2019}. We fix $\gamma=1,2,3$ in the fitting corresponding to loss with first, second, and third order kinetics. The best fit is obtained with $\gamma=2$, indicating that the loss mechanism is rate-limited by a two-body process with \mbox{$k_2$ = $(5.4\pm0.9)\times10^{-11}$\,cm$^{3}$\,s$^{-1}$}. This is consistent with loss mediated by the formation of two-body collision complexes.

\begin{figure}[t]
    \centering
			\includegraphics[width=0.49\textwidth]{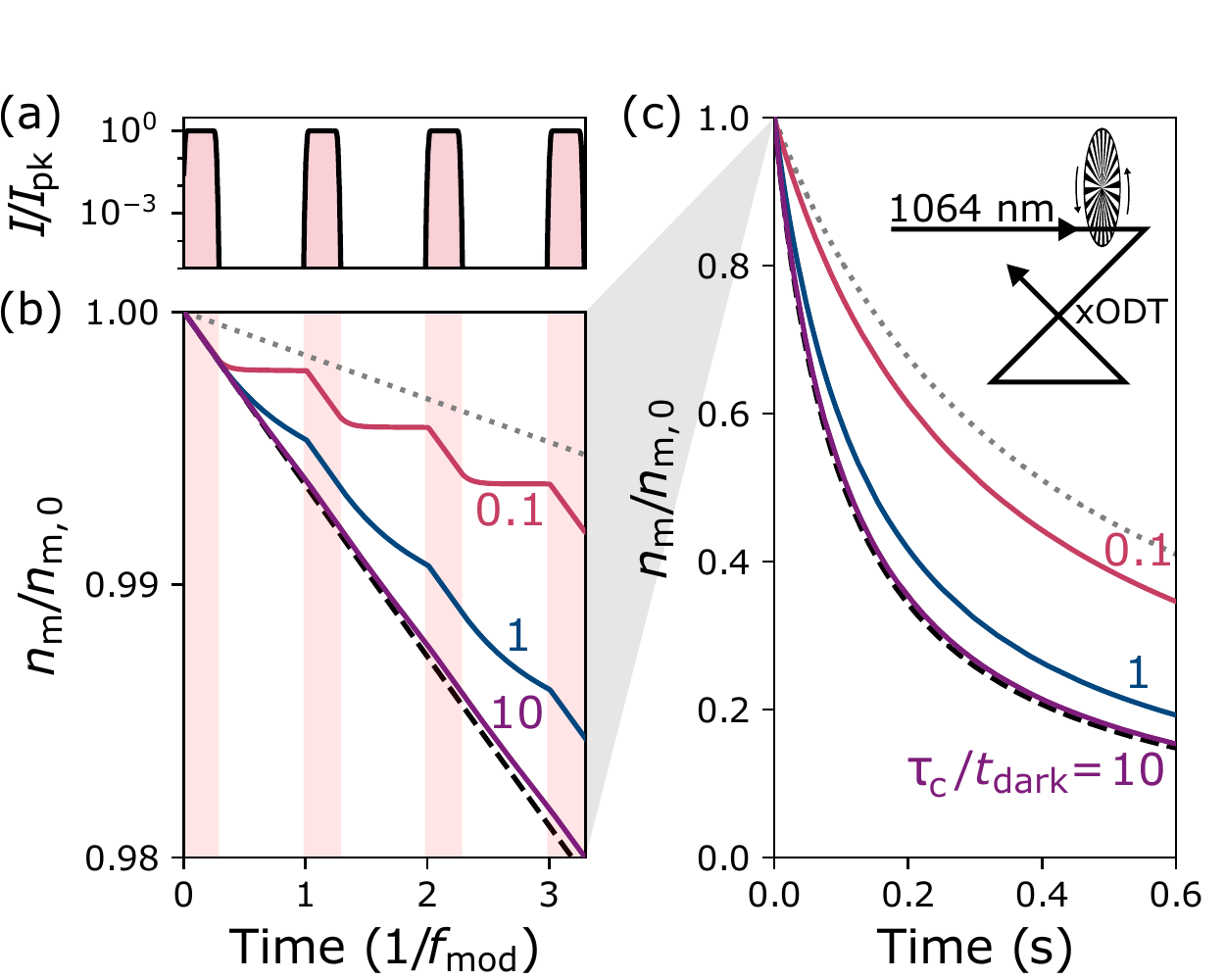}
		\vspace{-0.5cm}
    \caption{\label{fig:Introduction} Suppressing photoinduced two-body loss with an intensity-modulated trap. (a)~Intensity modulation of the trap, calculated from the beam waist measured at the position of the optical chopper. Solutions to Eq.~\ref{eq:RateEquations}, for CW and intensity-modulated traps are shown for (b) short and (c) long timescales. Molecule density $n_\mathrm{m}$ is plotted normalised to the starting density $n_\mathrm{m, 0}$. The dashed line indicates the loss from the CW trap. The coloured solid lines are for an intensity-modulated trap with $f_\mathrm{mod}=1.5$~kHz, 25\% duty cycle, and $\tau_\mathrm{c}/t_\mathrm{dark} = 0.1, 1, 10$. The dotted line indicates the expectation for a CW trap with a factor of 4 reduction in the two-body rate coefficient.}
\end{figure}

The average time for laser excitation of the complex is expected to be 2 to 3 orders of magnitude shorter than~$\tau_\mathrm{c}$~\cite{Christianen:2019}. Photoinduced loss of complexes may therefore be highly saturated, such that orders of magnitude reduction of the trap intensity is necessary to observe an intensity dependence. Reduction of the intensity by such a factor would catastrophically weaken a CW trap. Our use of square-wave modulation to generate a time-averaged trapping potential avoids this problem, as shown in Fig.~\ref{fig:Introduction}(a). When the intensity is high, the loss proceeds as in the CW trap. Note that to maintain trap depth, the peak intensity for the modulated trap is higher than for the equivalent CW trap, but as the complex loss is saturated, there is no change in the loss rate. When the trap light is off, however, there is no laser excitation of the complexes, and the loss is suppressed.

We model the rate of change of density of free molecules $\dot{n}_\mathrm{m}$ due to the formation, dissociation, and photoinduced removal of complexes as
\begin{align}
\begin{split}
    \dot{n}_\mathrm{m} &= -k_{2}n_\mathrm{m}^{2} + \frac{2}{\tau_\mathrm{c}}n_\mathrm{c}, \\
    \dot{n}_{c} &= +\frac{1}{2}k_{2}n_\mathrm{m}^{2} - \frac{1}{\tau_\mathrm{c}} n_\mathrm{c} - k_{\mathrm{l}} I(t) n_\mathrm{c}.
\label{eq:RateEquations}
\end{split}
\end{align}
Here, $n_\mathrm{c}$ is the density of complexes, and $k_\mathrm{l}$ is the complex-photon scattering rate per unit intensity $I$. We fix $k_2$ to the value measured for loss in the CW trap. Solutions to Eq.~\ref{eq:RateEquations} for CW and intensity-modulated traps are shown in Fig.~\ref{fig:Introduction}(b) and (c). Here we assume $k_{\mathrm{l}} I(t) \gg 1/\tau_\mathrm{c}$ when the trap light is on, as predicted~\cite{Christianen:2019}. The suppression of loss depends strongly upon the ratio of the complex lifetime to the dark time, $\tau_\mathrm{c} / t_\mathrm{dark}$; the dark time must be sufficiently long that a significant number of complexes can form and dissociate back to RbCs molecules between the destructive laser pulses.

\begin{figure}[t]
    \centering
			\includegraphics[width=0.49\textwidth]{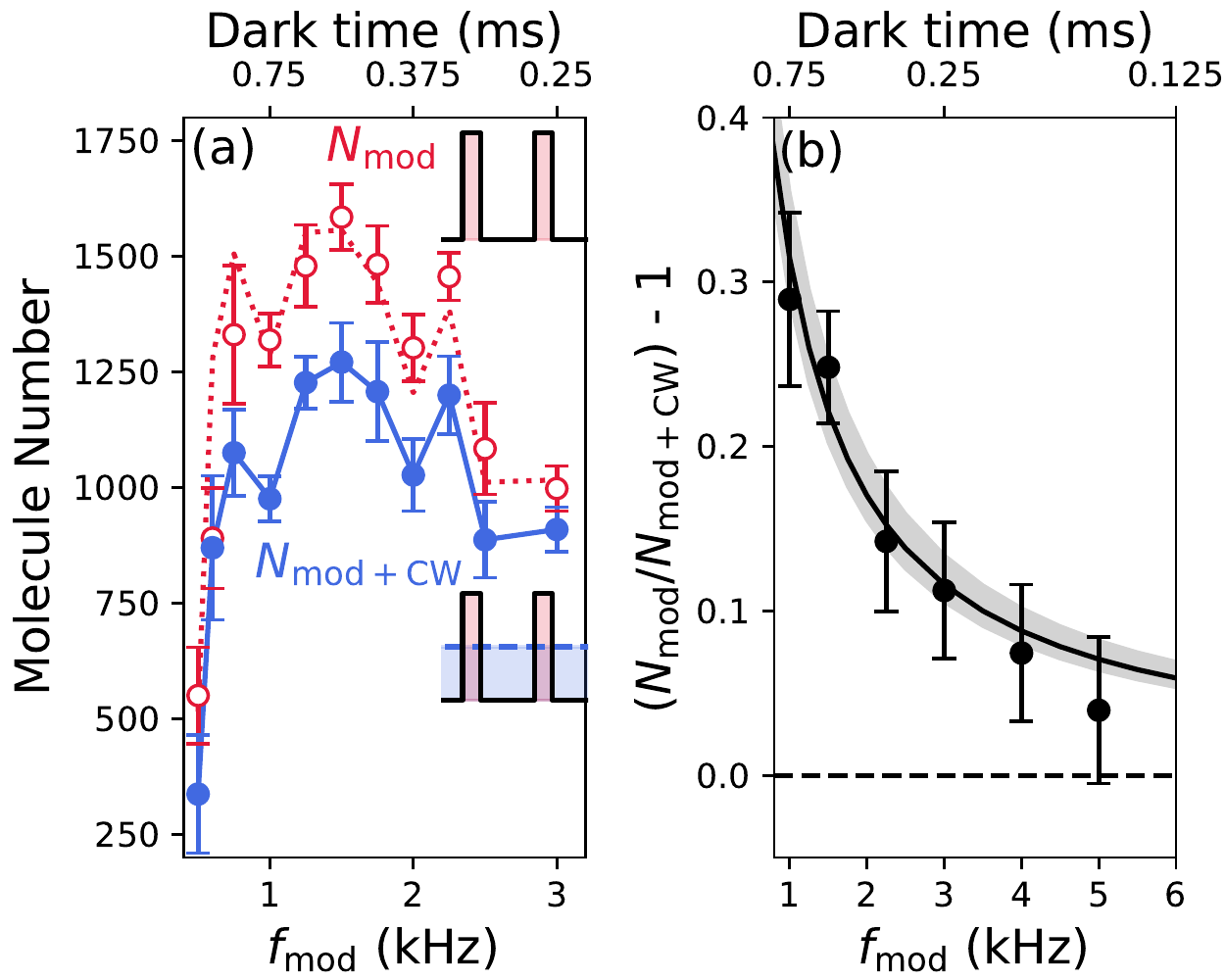}
    \vspace{-0.5cm}
    \caption{\label{fig:FrequencyDependence}{Frequency dependence of the loss suppression.} (a)~$N_{\mathrm{mod}}$ and $N_{\mathrm{mod+CW}}$ as a function of trap modulation frequency $f_{\mathrm{mod}}$. Each result shows the mean and standard error of 8 measurements. The solid blue line is a linear interpolation of the $N_\mathrm{mod+CW}$ results. The dotted red line is the interpolation for $N_\mathrm{mod+CW}$ multiplied by the best fit fractional difference shown in panel (b). (b)~Fractional difference in molecule number $N_\mathrm{mod}/N_\mathrm{mod+CW}-1$, where the dashed line indicates the expectation with no loss reduction. Each result is the ratio of mean for 50 measurements of $N_\mathrm{mod}$ and $N_\mathrm{mod+CW}$. An example histogram is shown in Fig.~\ref{fig:IntensityDependence}. The solid line is a best fit to the results, giving $\tau_\mathrm{c}=0.53\pm0.06$\,ms, with uncertainty in the fitting shown by the shaded region.}
    \label{fig:my_label}
\end{figure}

Modulating the intensity of the trap introduces an additional source of heating, which can lead to evaporative loss. To search for a reduction of photoinduced loss we therefore perform a comparative measurement, where the heating and thus evaporative loss are common. We measure the number of molecules remaining after 200\,ms in the trap with ($N_{\mathrm{mod+CW}}$) and without ($N_{\mathrm{mod}}$) additional CW 1550\,nm laser light, as illustrated inset in Fig.~\ref{fig:FrequencyDependence}(a). This light is derived from the trap used to prepare the Feshbach molecules. The total peak intensity of the CW light is $3.1(1)\times10^{2}$\,W\,cm$^{-2}$, whereas typical trap intensities are $\sim10^{4}$\,W\,cm$^{-2}$; this is sufficiently weak to not significantly affect the trap frequencies or trap depth, but high enough to continuously remove complexes from the trap, as we show in detail later. $N_\mathrm{mod}$ and $N_\mathrm{mod+CW}$ are shown as a function of trap modulation frequency in Fig.~\ref{fig:FrequencyDependence}(a), where each result is an average of 8 measurements. The trap frequencies experienced by the molecules during this measurement are $(\omega_x, \omega_y, \omega_z)/(2\pi)=(96(2), 160(3), 185(3))$\,Hz. We find the trap modulation introduces significant loss when $f_\mathrm{mod}<1$\,kHz; this is consistent with loss observed in time-averaged optical potentials resulting from a combination of parametric heating and not being fully in the time-averaged-trap regime~\cite{Schnelle:2008, Henderson:2009, Roberts:2014, Bell:2016}. Nevertheless, there is a broad range of modulation frequencies above 1\,kHz where $N_{\mathrm{mod}}$ is significantly greater than $N_{\mathrm{mod+CW}}$, indicating an observable reduction of the complex loss. 

To investigate the reduction in loss in more detail, we perform 50 interleaved measurements of $N_\mathrm{mod}$ and $N_\mathrm{mod+CW}$, and extract a mean and standard error from the resulting distributions. We characterise the loss reduction by $(N_\mathrm{mod}/N_\mathrm{mod+CW}) - 1$, which is shown as a function of $f_\mathrm{mod}$ in Fig.~\ref{fig:FrequencyDependence}(b). The solid line shows the best fit of Eq.~\ref{eq:RateEquations} to the results, again assuming $k_{\mathrm{l}} I(t) \gg 1/\tau_\mathrm{c}$ when the trap light is on. In this limit, the complex lifetime is the only free parameter, and we find $\tau_\mathrm{c}=0.53\pm0.06$\,ms. We estimate an additional systematic uncertainty of $\pm0.11$\,ms associated with our calibration of the density. We assume that the molecules remain at their initial temperature throughout, and neglect heating from the intensity modulation and two-body loss. However, from our model we estimate that a 25\% increase in the temperature over the course of the measurement would lead to a $\sim20$\% reduction in the fitted $\tau_\mathrm{c}$. 

Christianen~$et$~$al.$ predict a lifetime for the (RbCs)$_2$ complex of 0.253\,ms~\cite{Christianen:2019_DOS}. Their prediction is extrapolated from a detailed calculation for (NaK)$_2$ using approximate scaling laws. We note that the density of states increases strongly when moving from lighter to heavier systems, and that NaK is the lightest nonreactive heteronuclear bialkali molecule, while RbCs is the heaviest. A more accurate calculation for (RbCs)$_2$ will therefore be challenging~\cite{Groenenboom:PrivateCommunication}. Nevertheless, our measured $\tau_\mathrm{c}$ compares very favourably with the scaled prediction.

\begin{figure}[t]
    \centering
    \vspace{-0.25cm}
			\includegraphics[width=0.49\textwidth]{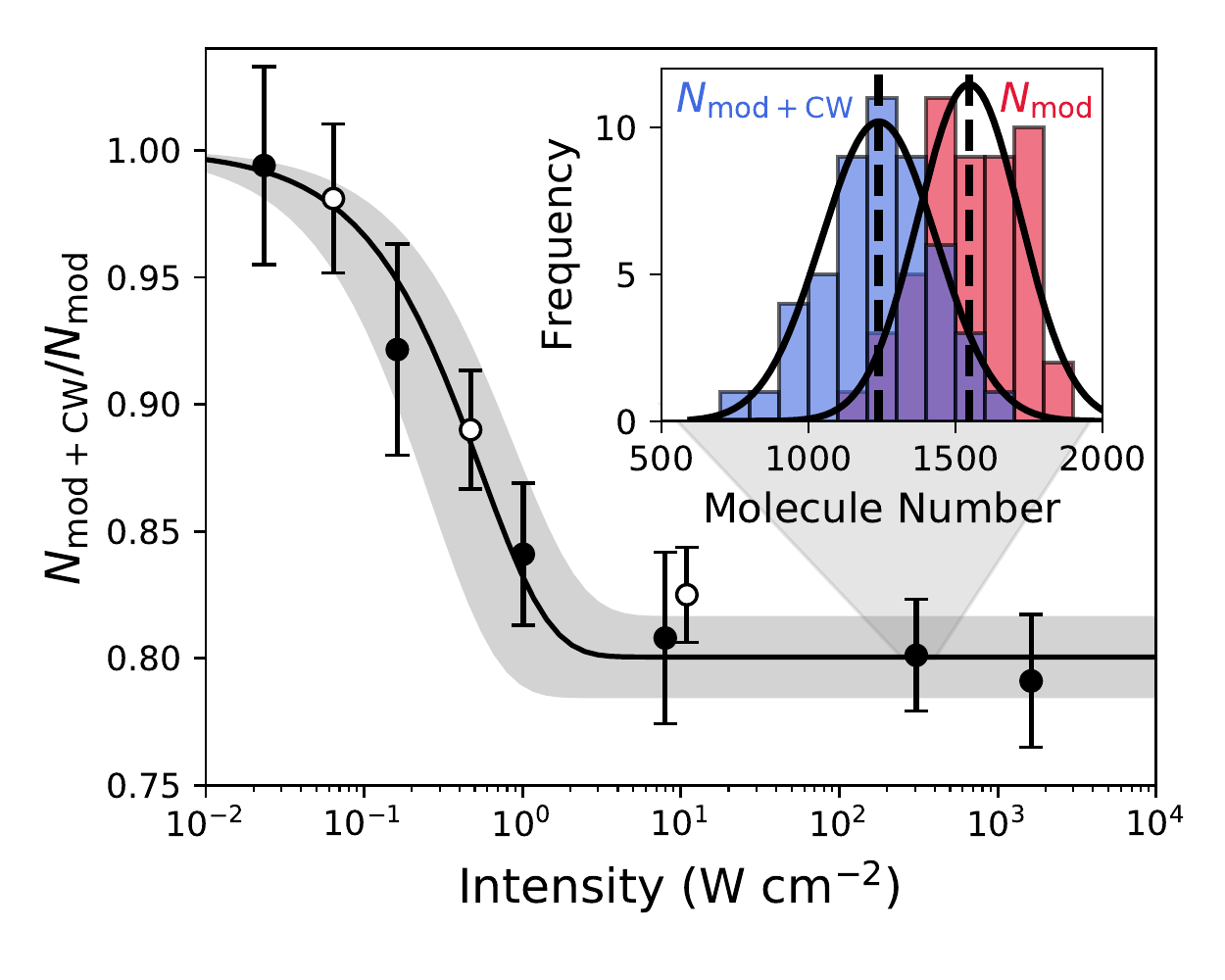}
	\vspace{-1cm}
    \caption{\label{fig:IntensityDependence}{Intensity dependence of the complex loss.} With the modulation frequency fixed at $f_{\mathrm{mod}}$ = 1.5 kHz, we observe the intensity dependence of the fractional change in molecule number $N_\mathrm{mod+CW}/N_\mathrm{mod}$. Filled (empty) circles indicate measurements with 1550\,nm (1064\,nm) light. Error bars indicate $1\sigma$ standard errors. The solid line shows a fit to the 1550~nm results, given by equation~\ref{eq:IntensityDependence}. We find $k_{\mathrm{l}}\tau_\mathrm{c}=2^{+2}_{-1}$~W$^{-1}$~cm$^{2}$. For each result, we perform 50 interleaved measurements of $N_{\mathrm{mod}}$ and $N_{\mathrm{mod+CW}}$. An example histogram is shown inset.}
\end{figure}

To investigate the complex-photon scattering rate, we measure the ratio $N_\mathrm{mod+CW}/N_\mathrm{mod}$ as a function of the CW laser intensity, keeping $f_\mathrm{mod}=1.5$\,kHz constant, as shown by the filled circles in Fig.~\ref{fig:IntensityDependence}. To access intensities below 10\,W\,cm$^{-2}$ we use a larger diameter 1550\,nm beam. We fit the intensity dependence as 
\begin{equation}
\frac{N_{\mathrm{mod+CW}}}{N_{\mathrm{mod}}} = (1-B)\exp{(-k_\mathrm{l}\tau_\mathrm{c} I)} + B,
\label{eq:IntensityDependence}
\end{equation}
where $B$ is the fraction of molecules remaining once the complex loss has been saturated. With $k_{\mathrm{l}}\tau_\mathrm{c}$ and $B$ as free parameters, we find ${k_{\mathrm{l}}\tau_\mathrm{c}=2^{+2}_{-1}}$~W$^{-1}$~cm$^{2}$. Combining this with our measurement of $\tau_\mathrm{c}$, we determine the intensity-normalised laser scattering rate ${k_\mathrm{l}=3^{+4}_{-2}\times10^{3}}$\,W$^{-1}$\,cm$^{2}$~s$^{-1}$. We have also performed experiments where the $1550$\,nm CW light is replaced with $1064$\,nm light, shown by the empty circles in Fig.~\ref{fig:IntensityDependence}; we observe no significant difference in behaviour between these two wavelengths.

The intensity required to achieve a 10\,$\upmu$K deep 1550\,nm CW optical trap is $5\times10^{3}$\,W\,cm$^{-2}$~\cite{Gregory:2017}. The average time for laser excitation of the complex at this intensity is $<100$\,ns. This is three orders of magnitude smaller than $\tau_\mathrm{c}$, confirming the prediction~\cite{Christianen:2019} that $k_{\mathrm{l}} I(t) \gg 1/\tau_\mathrm{c}$ and validating our earlier assumption. For a CW trap, the photoinduced loss is therefore highly saturated, and the depth would need to be reduced to $\sim$nK for an observable change in the loss rate. Christianen~$et$~$al.$ note that significant changes in the laser excitation rate may require increasing the wavelength of the light to $\sim10\,\upmu$m~\cite{Christianen:2019}. 
Alternatively, blue-detuned box-like traps may be used to reduce the photon scattering rate~\cite{Grimm:2000}. Achieving the factor of $\sim10^4$ reduction in the average intensity needed to suppress the loss in RbCs will be difficult, but this approach may work for molecules with shorter complex lifetimes. The use of a magnetic trap~\cite{Williams:2018} would avoid the photoinduced loss altogether and is an interesting prospect for molecules with an electronic magnetic moment, such as laser-cooled doublet molecules~\cite{Williams:2018} and bialkali triplet molecules~\cite{Lang:2008,Rvachov:2017}. Finally, there are numerous proposals for preventing molecular collisions reaching short range, thereby avoiding complex formation, by inducing repulsive interactions between colliding molecules using static electric fields~\cite{Gonzalez-Martinez:2017}, or microwave fields~\cite{Karman:2018, Lassabliere:2018}. 

Our use of an intensity-modulated trap to suppress photoinduced loss offers powerful insight into the lifetime of the collision complex, both in the dark and in the optical trap light. This gives an opportunity to benchmark theoretical methods as key experimental parameters are changed. For example, the number of energetically available open channels may affect the complex lifetime, and can be changed by using molecules in higher hyperfine or rotational states. Similarly, applying external fields, such as a DC electric field, may also affect the complex lifetime. Furthermore, a comparison of the complex lifetime across a range of molecule-molecule and atom-molecule systems will test the predicted scaling with the density of states. In addition, measuring the intensity dependence of the excitation enables the search for trap wavelengths which minimise the loss and may also offer a sensitive probe with which to search for scattering resonances.

In conclusion, we have demonstrated that complex-mediated photoinduced losses are the dominant source of loss in optically trapped samples of ground-state RbCs molecules. Our observations verify the mechanism proposed by Christianen~$et$~$al.$~\cite{Christianen:2019} to explain the two-body loss oberved in experiments using nonreactive bialkali molecules. We have shown that the loss may be partially suppressed by square-wave modulation of the trap intensity, such that the molecules spend 75\% of each modulation cycle in the dark. By varying the frequency of the modulation, we have measured the lifetime of the collision complex $\tau_\mathrm{c}=0.53\pm0.06$\,ms in the dark. We find the intensity-normalised laser excitation rate for the complex of $3^{+4}_{-2}\times10^{3}$\,W$^{-1}$\,cm$^{2}$\,s$^{-1}$ for a wavelength of 1550\,nm. For RbCs, this indicates that for typical trap intensities, the excitation is saturated by many orders of magnitude. Our approach offers an accessible new method to probe molecular collision complexes, benchmarking theoretical models and advancing the understanding of ultracold molecular collisions.

We acknowledge inspirational discussions with Martin Zeppenfeld, Tijs Karman, Gerrit Groenenboom, Arthur Christianen, and Jeremy Hutson. We thank Kevin Weatherill for the loan of the optical chopper. This work was supported by U.K. Engineering and Physical Sciences Research Council (EPSRC) Grants EP/P008275/1 and EP/P01058X/1. The data presented in this work are available at DOI:10.15128/r2j9602061h.

Following submission of this manuscript, we became aware of a related experimental work~\cite{Liu:2020}.

\bibliography{References}
\end{document}